\documentclass[graphics,floatfix,footinbib,tightenlines,nobibnotes,aps,prb,twocolumn]{revtex4-1}
\usepackage{amsmath}
\usepackage{amssymb}
\usepackage{graphicx}
\usepackage{comment}
\usepackage{soul,xcolor}
\usepackage{bbold}
\usepackage[colorlinks=true ,urlcolor=blue,urlbordercolor={0 1 1}]{hyperref}
\usepackage{tikz}
\usepackage{braket}
\usepackage[english]{babel}
\usepackage{graphicx}
\usepackage{lipsum}
\usepackage{xr}
\usepackage{ulem}
\newcommand{\up}{\uparrow}
\newcommand{\down}{\downarrow}

\renewcommand{\v}[1]{\mathbf{#1}} 

\newcommand{\be}{\begin{equation}}
\newcommand{\ba}{\begin{align}}
\newcommand{\ee}{\end{equation}}
\newcommand{\bea}{\begin{eqnarray}}
\newcommand{\eea}{\end{eqnarray}}
\newcommand{\beq}{\begin{equation}}
\newcommand{\eeq}{\end{equation}}
\newcommand{\beqn}{\begin{eqnarray}}
\newcommand{\eeqn}{\end{eqnarray}}

\newcommand{\Tr}{{\rm \, Tr\,}}

\newcommand{\bfk}{\mathbf{k}}
\newcommand{\bfq}{\mathbf{q}}

\newcommand{\moire}{moir\'e }

\newcommand{\tb}[1]{ \textcolor{blue} }

\begin{document}

\title{Theory of quantum anomalous Hall phases in  pentalayer rhombohedral graphene moir\'e structures}
 
\author{Zhihuan Dong}
\thanks{These authors contributed equally to this work.}
\author{Adarsh S. Patri}
\thanks{These authors contributed equally to this work.}
\author{T. Senthil}

\affiliation{Department of Physics, Massachusetts Institute of Technology, Massachusetts 02139, USA}
 
\date{\today}
\begin{abstract}
Remarkable recent experiments on the moir\'e structure formed by pentalayer rhombohedral graphene aligned with a hexagonal Boron-Nitride substrate report the discovery of a zero field fractional quantum hall effect. These ``(Fractional) Quantum Anomalous Hall"  ((F)QAH) phases occur for one sign of a perpendicular displacement field, and correspond, experimentally, to full or partial filling of a valley polarized Chern-$1$ band.  Such a band is absent in the non-interacting band structure. Here we show that electron-electron interactions play a crucial role, and present microscopic theoretical calculations demonstrating the emergence of a nearly flat, isolated, Chern-$1$ band and FQAH phases in this system. We also study the four and six-layer analogs and identify parameters where a nearly flat isolated Chern-$1$ band emerges which may be suitable to host FQAH physics.

\end{abstract}
\maketitle
Recent experiments have observed\cite{cai2023signatures,zeng2023integer,park2023observation,xu2023observation,lu2023fractional}  the fractional quantum Hall effect in systems that microscopically are time reversal invariant. The corresponding states of matter -- known as Fractional Quantum Anomalous Hall (FQAH) phases -- break time reversal spontaneously. FQAH states were first evidenced in twisted MoTe$_2$ moir\'e heterostructures through measurements of the location of the many body gap \cite{cai2023signatures,zeng2023integer} as a function of density and magnetic field, and subsequently, in transport\cite{park2023observation,xu2023observation}. 

Very recently, many FQAH states have been discovered\cite{lu2023fractional} in pentalayer rhombohedral graphene aligned with a hexagonal Boron-Nitride (hBN)  substrate. We show, through microscopic calculations, in this paper that these arise due to a novel mechanism where Coulomb interactions stabilize a nearly flat isolated Chern band which then sets the stage for the appearance of FQAH physics. 

 Rhombohedral graphene refers to a structure where different layers of graphene are stacked together in the ABC pattern shown in Fig.~\ref{fig_continuum_penta}(a). Without the hBN alignment,  a first-cut description of the band structure, when there are a total of $n$ such layers, is that valence and conduction bands have a very flat $n$th order band touching (with a $k^n$ dispersion where $k$ is the wavenumber) in each of the two valleys. A more accurate band structure, shown in Fig.~\ref{fig_continuum_penta}(b) that includes the effects of trigonal warping modifies the precise high-order band touching. The flatness of the bands near neutrality enhances the effects of Coulomb interaction, as evidenced by the discovery\cite{zhou2021half,zhou2021superconductivity,zhou2022isospin,de2022cascade,zhang2023enhanced,Han2023,liu2023interaction}, in the last few years, of interaction-driven many-body states in $n = 2, 3, 4, 5$ rhombohedral graphene. We will denote $n$-layer rhombohedral graphene as R$n$G below; if it is further aligned with hBN on one side we denote it as R$n$G/hBN.

\begin{figure}
    \centering
\includegraphics[width=0.49\textwidth] {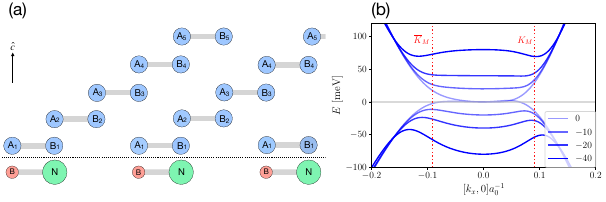}
    \caption{
    (a) Schematic of pentalayer rhombohedral graphene (R5G), with/without aligned hBN on layer 1. Carbon atoms highlighted in blue; Boron atoms in red; and Nitrogen atoms in green.
    The alignment is chosen here so that $A_1$ ($B_1$) aligns with Boron (Nitrogen).
    (b) Continuum R5G dispersion (without hBN alignment) along the $k_x$ axis for increasing interlayer potential difference $u_d$ [meV] highlighted by darker shades of blue. The vertical dashed lines indicate the location of the \moire $K$ points when hBN is aligned with pentalayer graphene.
    }
    \label{fig_continuum_penta}
\end{figure}

\begin{figure}
    \centering
\includegraphics[width=0.49\textwidth] 
{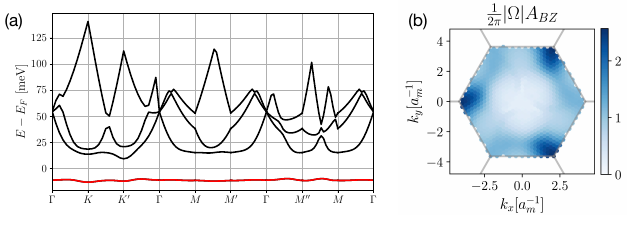}
    \caption{    (a) Hartree-Fock band structure of R5G/hBN for a single valley with an inter-layer potential difference $u_d = -36$ meV, and relative dielectric constant $\epsilon = 8$ and (b) Berry curvature $\Omega$ multiplied by the \moire Brillouin zone area $A_{BZ}$. The active conduction band (highlighted in red) is flat (bandwidth of 4 meV), well separated from the remote conduction bands above (direct gap of 19 meV and global gap of 17 meV), and carries a Chern number $|C| = 1$.
    Momentum mesh of 30$\times$30, and keeping the four lowest conduction bands.
    }
    \label{hf_ud_m36_eps_8}
\end{figure}

In RnG/hBN the slight lattice mismatch between graphene and hBN produces a moir\'e pattern which strongly modifies the electronic band structure\cite{chen2019evidence}. Theoretical study\cite{zhang2019nearly} of R3G/hBN showed that the moir\'e bands are nearly flat. Further, the band structure is very sensitive\cite{ zhang2019nearly} to a perpendicular displacement field $D$. For one sign of the displacement field, the bands are topological with valley Chern number $\pm 3$ while for the other sign, they have Chern number $0$. The topological side occurs when the active change carriers are driven away from the aligned layer. These features lead to the proposal\cite{zhang2019nearly} that these systems will, for one sign of $D$, be good platforms to obtain (fractional) quantum anomalous Hall effects with the requisite time-reversal breaking realized through valley (and spin) polarization. Indeed, experiments\cite{chen2020tunable} on  R3G/hBN at total filling $\nu_T = 1$ of the valence band observe an integer quantum anomalous Hall state with a Chern number $2$. The change of Chern number to $2$ from the naive band theory expectation of $3$ was explained\cite{chen2020tunable} through a Hartree-Fock calculation that includes mixing with the next remote band. 

Generalized to  R$n$G/hBN, the naive expectation is that for one sign of $D$, the bands will have a valley Chern number $\pm n$, and hence might show an integer quantum anomalous Hall effect with $\sigma_{xy} = \pm \frac{ne^2}{h}$ at total odd-integer filling of the \moire bands. However the experiments of Ref.~\onlinecite{lu2023fractional} on R5G/hBN see an integer quantum Hall effect with $\sigma_{xy} = \pm \frac{e^2}{h}$ at $\nu_T = 1$ of the conduction band. Further, the FQAH states observed at fractional filling ($\nu_T = 2/3, 3/5,...$) are what might be expected at partial fractional filling of a Chern-$1$ band.  The theoretical non-interacting band structure (at the large displacement field energies corresponding to those in Ref.~\onlinecite{lu2023fractional}) is shown in Fig.~\ref{fig3_non_int_moire}, and does not even have an isolated conduction band, let alone a Chern number. Thus a task for theory is to first explain the emergence of an isolated Chern-$1$ band in the R5G/hBN system and next to demonstrate that this Chern band hosts FQAH phases.   

Here, we present a Hartree-Fock treatment of the band structure in the presence of both an aligned hBN substrate as well as a perpendicular displacement field. We identify a range of displacement fields where there is an interaction-induced  first conduction band (see Fig.~\ref{hf_ud_m36_eps_8}) that is isolated from other bands, has a small bandwidth, and a net Chern number of $\pm 1$ in each of the two graphene valleys.  We provide evidence of valley polarization consistent with experiments. At partial rational fillings of the valley polarized bands,  we then show using exact diagonalization studies that FQAH phases develop. 
Our results provide a microscopic basis for the observation of FQAH physics in R5G/hBN.
We also explore the renormalized band structure of R$n$G/hBN, for $n = 4, 6$ finding parameters with nearly flat Chern-$1$ bands that are well separated from other bands. Thus R4G/hBN and R6G/hBN may also be good platforms for FQAH physics. 

The importance of electronic interactions in obtaining the nearly flat isolated Chern band in RnG/hBN makes them unique compared to the theory of previously explored platforms \cite{abouelkomsan2020particle,ledwith2020fractional,repellin2020chern,wilhelm2021interplay,xie2021fractional,wu2019topological,yu2020giant,devakul2021magic,li2021spontaneous,crepel2023fci,wang2023fractional,reddy2023fractional}  or of toy models\cite{sun2011nearly,sheng2011fractional,neupert2011fractional,wang2011fractional,tang2011high,regnault2011fractional,neupert2011fractionalb,bergholtz2013topological,parameswaran2013fractional}.

\textit{Model of R5G/hBN.---}
In  R5G, for a given spin and valley, there are 2 sublattice degrees of freedom per layer, leading to a description of the non-interacting physics in terms of a 10-band model. A perpendicular displacement field implies a potential difference between adjacent layers (denoted $u_d$). The resulting tight-binding model is described in Supplementary Materials (SM) \textcolor{red}{I}. At low energies near the charge neutrality point, the important degrees of freedom lie in one sublattice (say the A sublattice) in the bottom layer and in the opposite B sublattice in the top layer. A displacement field tends to drive electrons preferentially to either the top or bottom layer depending on its sign.

The R5G is subjected to an underlying moir{\'e} potential through the (near) alignment of one end, which we take to be the bottom layer, with a hBN substrate. We choose an alignment angle of $0.77^{\circ}$ to match the angle quoted in Ref. \onlinecite{lu2023fractional}. 
The moir{\'e} potential is modeled as,
\begin{align}
\label{eq_moire_pot_main}
H_M = \sum_{\mathbf{G}} c_1 ^{\dag} ( \mathbf{k} + \mathbf{G} ) V_M (\mathbf{G}) c_1 (\mathbf{k}),
\end{align}
where $V_M(\mathbf{G})$ is defined in SM \textcolor{red}{I},  and $\mathbf{G}$ are the reciprocal lattice vectors of the \moire superlattice.

 To discuss many-body physics, we take a dual gate-screened Coulomb interaction between the low-energy electrons,
\begin{align}
\label{eq_screened_coulomb}
H_{\text{C}} = \frac{1}{2 A} \sum_{\bfk, \bfk', \bfq} \sum_{\mu, \nu} V^{\text{sc}}_C (\bfq) c^{\dag}_{\bfk + \bfq; \mu} c^{\dag}_{\bfk' - \bfq; \nu}  c_{\bfk' ; \nu}  c_{\bfk; \mu} , 
\end{align}
where $A$ is the area of the system, $\mu, \nu$ denotes a compact sublattice-valley-flavor index, $V^{\text{sc}}_C (\bfq)= \frac{e^2}{2 \epsilon_0 \epsilon q} \tanh(q d_s)$ is the screened potential, $\epsilon$ is the effective dielectric constant, and $d_s$ is the distance between the metallic gates and the top and bottom layers (taken to be 30nm).


We begin with the non-interacting band structure in the moir\'e Brillouin zone.
Figure~\ref{fig_continuum_penta}(b) shows the continuum dispersion for one valley of isolated  R5G for increasing displacement field energies $u_d<0$. {\color{black} The sign of $u_d$ is such that electrons are driven away from the aligned hBN.} 
For $u_d=0$, there is a band-touching at the $\Gamma$-point. 
As the magnitude of the displacement field energy is increased, a band gap develops in conjunction with a flattening of the conduction band.

\begin{figure}
    \centering
\includegraphics[width=0.48\textwidth]{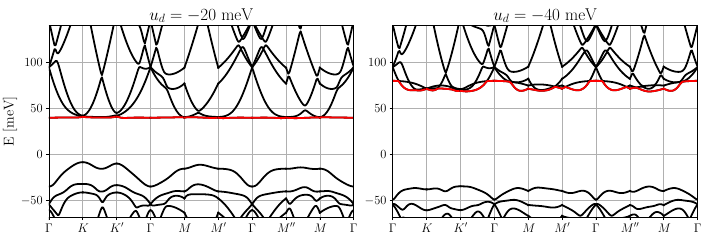}
\caption{Non-interacting bandstructure for R5G/hBN for displacement field energies $u_d = -20, -40$ meV. There is a wide gap between valence and conduction bands.}
\label{fig3_non_int_moire}
\end{figure}

When subjected to an aligned \moire potential, the continuum band reconstructs into the \moire Brillouin zone.
The \moire $K$ point is at $\approx 0.1/a_0$; as seen in Fig.~\ref{fig_continuum_penta}(b) this corresponds to the low-energy conduction \moire bands being formed predominantly by the ``flat'' region of the conduction continuum band.
Figure~\ref{fig3_non_int_moire} shows the \moire bandstructure for displacement fields of $u_d = -20, -40$ meV.
With increasing displacement field, the active conduction band (highlighted in red) separates away from the valence bands but  
eventually collides with the higher `remote' conduction bands for $u_d \gtrsim -40$ meV. Thus for a large displacement field, we expect that interaction effects will mix together various conduction bands, and lead to a renormalized band sructure with modified Bloch functions, and possibly, open up a band gap. 

\begin{figure}
    \centering
    \includegraphics[width=0.5\textwidth]{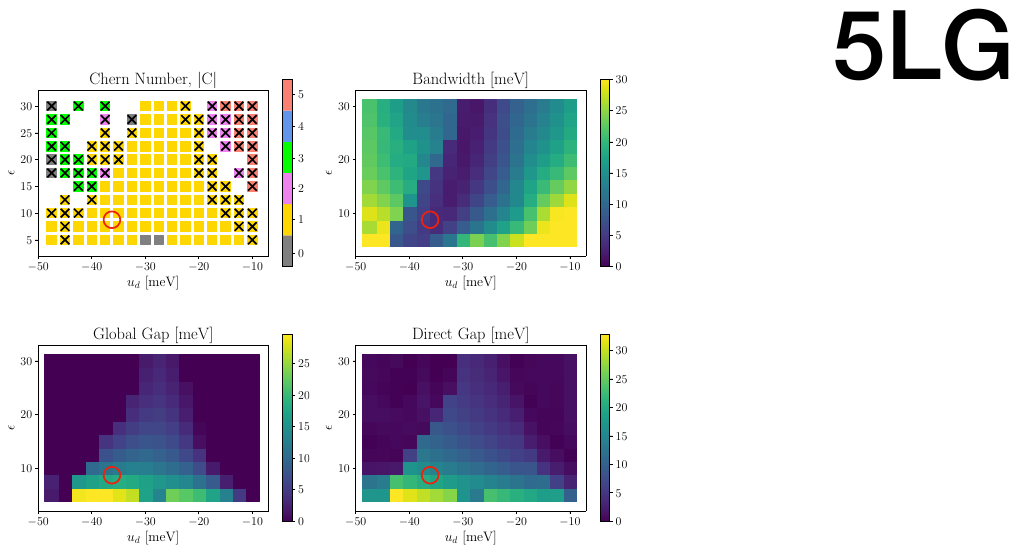}
    \caption{Hartree-Fock bands indicating Chern number and bandwidth of the active band, global band gap, and direct band gap to nearest conduction band for different displacement field energies $u_d$ and dielectric constant $\epsilon$.
    The Chern number is well defined when the direct band gap is non-zero ($\geq0.5$meV).
    The `crossed' out yellow boxes in the Chern number indicate cases where the global bandgap is zero ($<0.5$meV), while the direct band gap is still non-zero (i.e. there is an indirect band gap -- the system is metallic).
    Phase diagrams are for a mesh of 27$\times$27 and constructed using 19 \moire Brillouin zones, and generated using at least five distinct initial mean-field ansatzes for a given parameter point.
    The red-circled region is the parameter chosen for the ED.
    }
    \label{fig_hf_phase_diagram}
    \end{figure}

 

\textit{Hartree-Fock interaction effects for R5G/hBN.---}
To deal with these effects, we perform a Hartree-Fock calculation (see SM \textcolor{red}{II}) of the band structure in a single valley (and ignoring electron spin) keeping the first three conduction bands.  
We present in Fig.~\ref{hf_ud_m36_eps_8} the renormalized bandstructure and the Berry curvature $\Omega$. 
The band is very flat (bandwidth of $\approx 4$ meV), with a global gap to the neighboring conduction band of $ \approx 17$ meV.
Moreover, the Berry curvature has redistributed to make the band have $|C| = 1$. We emphasize that both the band gap and the Chern number are interaction-induced and are absent in the free electron band structure. 


Figure~\ref{fig_hf_phase_diagram} shows the Hartree-Fock band structure `phase diagram'  as the displacement field and dielectric constant are swept over a range of parameters.
There is a stripe in parameter space of obtaining $|C|=1$ band that is conducive to forming a correlated state i.e. has a narrow bandwidth and a well-separated gap from the other bands.
Indeed, there is an optimal window (i.e. minimum bandwidth and maximum gap) for $|u_d| \approx 35 - 40$ meV and $\epsilon \approx 5-10$.
The prevalence of the Chern band (over the topologically trivial band) can be heuristically understood as a consequence of the Wannier obstruction of topological bands:
the prohibition of exponentially localized real-space wavefunctions leads to a more evenly distributed charge density, thereby reducing the Coulomb repulsion.

To further characterize the renormalized band structure in Fig.~\ref{fig_36_8_occup_metric}, we show the quantum metric $g_{ij} $, and the quantity $T = \frac{A_{BZ}}{2\pi} \int_{BZ} d^2k~ {\text{tr}}(g) - |\Omega|$ where $A_{BZ}$ is the area of the Brillouin zone. For an `ideal' Chern band\cite{roy2014band,parameswaran2013fractional,ledwith2022vortexability} $T = 0$. 
Intriguingly, the $|C| = 1$ conduction band has an average trace violation of $\langle T \rangle \approx 0.9$; for comparison,  for the first Landau Level this violation is $2$.

\begin{figure}
    \centering
    \includegraphics[width=0.4\textwidth]{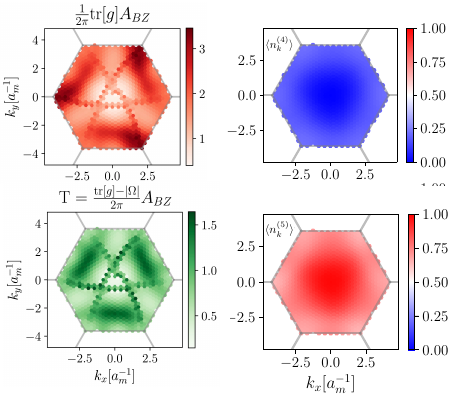}
 \caption{Hartree-Fock (HF) determined quantum metric, and trace condition, accompanied by the layer-4,5 occupation number. The total weight on the other 3 layers combined is $\approx 5 \%$.
    The mean and variance of the trace condition over the BZ is $\langle T \rangle = 0.9$ and var$(T) = 0.09$, respectively.
    }
    \label{fig_36_8_occup_metric}
\end{figure}

 In Fig.~\ref{fig_36_8_occup_metric}, we also show the occupation numbers in the top layer and the layer directly underneath it. At these large displacement fields, the electrons mostly reside just in the top layer (with some small charge density in the adjacent layer below). 
 
 R5G/hBN has four flavor degrees of freedom (spin and valley). Spontaneous valley polarization breaks time-reversal symmetry, and 
 enables the electrons to have a net non-zero Chern number. If, in addition, the spin is also polarized, then the many-body state at total filling $\nu_T = 1$ will be an Integer Quantum Anomalous Hall (IQAH) state. This is the mechanism\cite{zhang2019nearly,bultinck2020mechanism,zhang2019twisted,repellin2020ferromagnetism} for the ferromagnetism\cite{sharpe2019emergent} and the IQAH effect\cite{chen2020tunable,serlin2020intrinsic,chen2021electrically,polshyn2020electrical} observed in graphene moir\'e structures. 
 
 We present the Hartree-Fock bandstructure including both spin and valley degrees of freedom in Fig.~\ref{fig_valley_polarized_13}. 
 The lowest conduction band is fully spin-valley-polarized.
 However the fate of the spin deserves more discussion. 
In the main text, we assume full spin polarization and relax this assumption in SM \textcolor{red}{IV}.  
\begin{figure}
    \centering
\includegraphics[width = 0.45\textwidth]{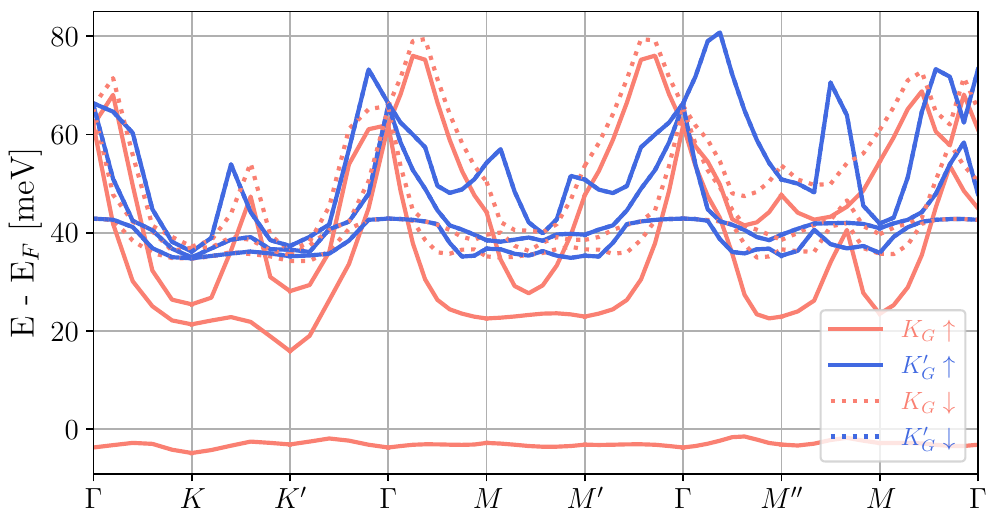}
    \caption{Spin-Valley polarization of Hartree-Fock band structure for displacement field $u_d = -36$ meV, and relative dielectric constant $\epsilon = 8$. 
    The colors label two valleys, and the solid/dotted lines indicate the spin orientation $\uparrow$/$\downarrow$.
    The active conduction band is polarized in the monolayer graphene valley $\{K_G, \uparrow \}$ throughout the Brillouin zone.
    The spin-valley polarization is spontaneously chosen (with a non-spontaneously chosen spin-polarization degeneracy remaining in the bandstructure).
    This calculation is performed on a momentum mesh of 15$\times$15 and keeping the three lowest conduction bands.
    }
\label{fig_valley_polarized_13}
\end{figure}

\textit{Evidence of FQAH from exact diagonalization.---}
We now project (see SM \textcolor{red}{II})  the Coulomb interaction to the active conduction Chern band and demonstrate that FQAH states emerge within an exact diagonalization study.  We study the spectrum of the resulting many-body Hamiltonian in a torus geometry, as shown in Fig.~\ref{twothirds}(a) and Fig.~\textcolor{red}{A.2}. We focus on band filling $2/3$ here; data at other fillings is in SM~\textcolor{red}{III}. At $2/3$ filling, there are a set of 3 nearly degenerate states that are well-separated (and lower) in energy than other states, consistent with a FQAH state.  

Next, we study the evolution of the spectrum upon inserting a single flux quantum through one cycle of the torus. In a FQAH state, this flux threading is equivalent to generating a quasi-hole and quasi-particle pair, moving one of them around one non-contractable loop of the torus, and annihilating the pair. This connects topologically distinct ground states on the torus. 
On a system with $N_x\times N_y$ geometry and at filling $\nu$, flux threading of $\Delta \Phi$ leads to a many-body momentum shift of
\be
\label{dk}
    \Delta k_i = \frac{\Delta\Phi_i}{L_i} L_xL_y\nu.
\ee
The momentum shift for the geometry/filling in Fig.~\ref{twothirds}(a) is consistent with this expectation. 

The spectral flow under unit flux insertion is shown in Fig.~\ref{twothirds}(b). 
The 3 nearly degenerate ground states are permuted under the flux insertion, consistent with the FQAH state. Furthermore, these states remain well separated from the excited states at all values of the flux. 

\begin{figure}
    \centering
\includegraphics[width=0.48\textwidth]{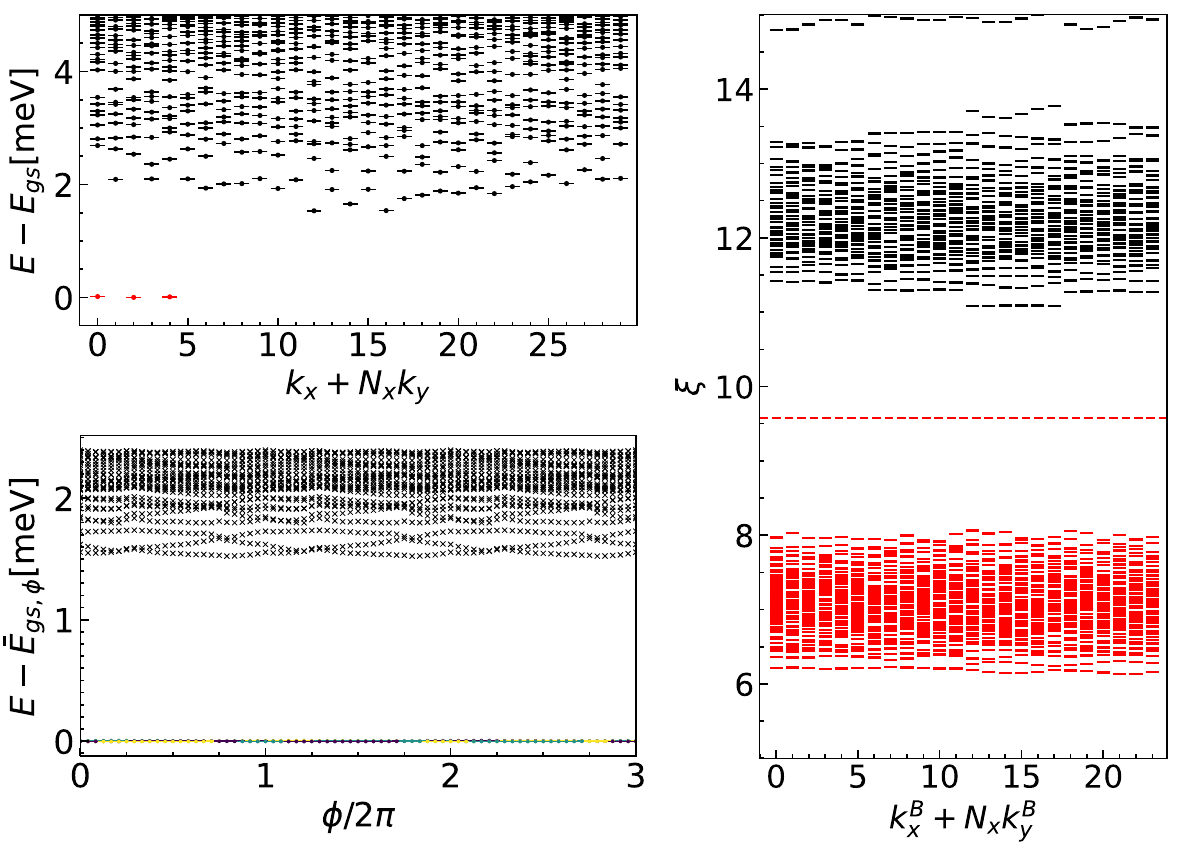}
    \caption{Evidence for FQAH at $\nu=2/3$ from exact diagonalization. (a) Low energy spectrum of a spin-valley polarized system with $N_x \times N_y = 6\times5$. (b) Spectral flow under flux threading along $x$ direction on $6\times4$ system. $\bar{E}_{gs}$ is the average energy for topologically degenerate ground states. The colors are identified by the momentum sector corresponding to the eigenstates. (c) Particle entanglement spectrum (PES) for the particle-hole conjugate of $\nu=2/3$ states. With a subsystem containing $N_B=3$ holes on a $6\times4$ lattice, we find the total number of states below the gap is 1088, consistent with the expectation\cite{regnault2011fractional} for FQAH. These results are obtained at $\epsilon=8$, $D=-36$meV. }
    \label{twothirds}
\end{figure}

In Fig.~\ref{twothirds}(c), we demonstrate the particle entanglement spectrum(PES)\cite{regnault2011fractional, sterdyniak2011pes}, which encodes the information of the quasihole Hilbert space dimension. For this purpose, we partition the system into subsystems A and B, with $N_A$ and $N_B$ electrons and momentum  $k_A$ and $k_B$ respectively. The PES $\xi_i$ is associated with the eigenvalue $\lambda_i$ of reduced density matrix $\rho_B = \Tr_A \rho$ through $\xi_i = -\log \lambda_i$. 
For $\nu=2/3$, the ground states are three-fold degenerate, and we take $\rho=\frac{1}{3}\left( \ket{\tilde{\Psi}_1}\bra{\tilde{\Psi}_1}+\ket{\tilde{\Psi}_2}\bra{\tilde{\Psi}_2}+\ket{\tilde{\Psi}_3}\bra{\tilde{\Psi}_3}\right)$, where $\ket{\tilde{\Psi}_i} = P \ket{\Psi_{i}}$ are the particle-hole conjugate of topologically degenerate ground states. The counting of states agrees exactly with that expected for the particle-hole conjugate of the $1/3$ Laughlin state (see also Fig.~\textcolor{red}{A.4}(c)). Taken together our calculations strongly support the identification of the state at $2/3$ filling with an FQAH state. Other gapped states such as ones with Charge Density Wave order are not compatible with these results.

 In SM \textcolor{red}{IV}, we present calculations that do not assume spin polarization, and find a rich and non-monotonic behavior as the filling is varied.



\textit{Discussion.---} 
An interesting question is the role of the \moire potential in the emergence of the isolated Chern-$1$ band.  At $\nu = 1$, we checked that such a band emerges in Hartree-Fock even in the absence of \moire (SM~\textcolor{red}{VI}), consistent with Refs. \onlinecite{zhou2023fractional,dong2023anomalous}. \textcolor{black}{This suggests a picture of the $\nu = 1$ state as a `Hall crystal' \cite{PhysRevB.39.8525} weakly pinned by the \moire potential that should be examined beyond the Hartree-Fock approximation. As we emphasize elsewhere, it is likely that the role of \moire is to stabilize the Hall crystal relative to a correlated Fermi liquid; our calculations should be viewed as a description of this moir\'e-enabled Hall crystal\cite{dong2024stability}. }  Finally, we discuss the prospect of FQAH physics in other R$n$G/hBN systems. In SM 
~\textcolor{red}{VII} and \textcolor{red}{VIII}, we study the renormalized band structures of R4G/hBN and R6G/hBN for a single valley and spin within Hartree-Fock theory as a function of displacement field energy and dielectric constant. For concreteness, we take the same alignment angle with hBN as in our studies of R5G/hBN. For both R4G/hBN and R6G/hBN, we find parameter regimes where there is a well-isolated nearly flat Chern-$1$ band, and with quantum geometry comparable to that in R5G/hBN. This agrees with the expectation that since at large displacement fields, only the top two layers away from the hBN have any occupation at all, the total number of layers does not matter much. However, the optimal range of the displacement field decreases with increasing $n$, consistent with the expectation that the bands are flatter for larger $n$. 
In all these systems, the physics is sensitive to the alignment angle with hBN. We thus anticipate a rich and interesting phase diagram to be unearthed in R$n$G/hBN as more of the parameter space is explored experimentally.

After this work was posted in the arXiv, two papers\cite{zhou2023fractional,dong2023anomalous} with closely related, and consistent,  results also appeared. 

\acknowledgements
We thank Long Ju, Tonghang Han, and Zhengguang Lu for discussions of their experimental data, C\'ecile Repellin for sharing her insights on exact diagonalization studies of quantum hall states, and Ya-Hui Zhang for many inspiring conversations and previous collaborations. 
TS was supported by NSF grant DMR-2206305, and partially through a Simons Investigator Award from the Simons Foundation. This work was also partly supported by the Simons Collaboration on Ultra-Quantum Matter, which is a grant from the Simons Foundation (Grant No. 651446, T.S.). 
The authors acknowledge the MIT SuperCloud and Lincoln Laboratory Supercomputing Center for providing HPC resources that have contributed to the research results reported within this manuscript.

\maketitle
\bibliographystyle{apsrev4-1}
\bibliography{FQAH}

\appendix

\renewcommand\thefigure{A.\arabic{figure}}
\renewcommand\theequation{A.\arabic{equation}}
\setcounter{figure}{0}

\section{Pentalayer graphene tight-binding and \moire parameters}
\label{app_penta_params}
For a given spin and valley ($\pm$), the tight-binding Hamiltonian is given by (adapting the model for rhombohedral-stacked trilayer graphene \cite{macdonald_trilayer_2010}),
\begin{widetext}
\begin{align}
\label{full_hamiltonian_penta}
H_0 = 
\begin{pmatrix}
2 u_d & v_0^{\dag} & v_4^{\dag} & v_3 & 0 & \frac{\gamma_2}{2} & 0 & 0 & 0 & 0\\
v_0 & 2 u_d + \delta & \gamma_1 & v_4^{\dag} & 0 & 0 & 0 & 0 & 0 & 0 \\
v_4 & \gamma_1 & u_d + u_a & v_0^\dag & v_4^\dag & v_3 & 0 & \frac{\gamma_2}{2} & 0 & 0 \\
v_3^{\dag} & v_4 & v_0 & u_d + u_a & \gamma_1 & v_4^{\dag} & 0 & 0 & 0 & 0 \\
0 & 0 & v_4 & \gamma_1 & u_a & v_0^{\dag} & v_4^{\dag} & v_3 & 0 & \frac{\gamma_2}{2} \\
\frac{\gamma_2}{2} & 0 & v_3^{\dag} & v_4  & v_0  & u_a & \gamma_1 & v_4 ^{\dag} & 0 & 0 \\
0 & 0 & 0 & 0 & v_4 & \gamma_1 & -u_d + u_a & v_0^{\dag} & v_4^{\dag} & v_3 \\
0 &0 & \frac{\gamma_2}{2} & 0 & v_3^{\dag} & v_4 & v_0 & -u_d + u_a & \gamma_1 & v_4^{\dag} \\
0 & 0 & 0 & 0 & 0 & 0 & v_4 & \gamma_1 & -2u_d + \delta & v_0^{\dag}\\
0 & 0 & 0 & 0 & \frac{\gamma_2}{2} & 0 & v_3^{\dag} & v_4 & v_0 & -2u_d  
\end{pmatrix},
\end{align}
\end{widetext}
where we employ the basis of $(A_1, B_1, A_2, B_2, A_3, B_3, A_4, B_4, A_5, B_5)$.
We have used the compact notation of $v_i \equiv v_l \Pi = \frac{\sqrt{3}}{2} t _l (\pm k_x + i k_y)$ for the monolayer graphene form factors, where $k_{x,y}$ are small momenta expanded about the valley of interest, and the layer is denoted by $l$.

The on-site potentials include the effects of the displacement energy and are incorporated by adding a potential difference $u_d$ between adjacent layers. 
We choose the following tight-binding parameters given in Table \ref{tab:tight_binding}, where the on-site potentials are in agreement with those of rhombohedral stacked tetralayer graphene \cite{parkjeil2023}.
We take the bare $\gamma_0$ hopping obtained from density functional theory (DFT), since we are implementing the effects of interactions through both Hartree-Fock and within the exact diagonalization (ED) framework.
\begin{table}[h!]
\centering
\begin{tabular}{|c|c|c|c|c|c|c|}
\hline
$\gamma_0$ & $\gamma_1$ & $\gamma_2$ & $\gamma_3$ & $\gamma_4$ & $\delta$ & $u_a$ \\
\hline
$2600$ & $356.1$ & $-15$ & $-293$ & $-144$ & 12.2 & $-16.4$ \\
\hline
\end{tabular}
\caption{Tight-binding parameters for rhombohedral-stacked pentalayer graphene (in meV).}
\label{tab:tight_binding}
\end{table}

A \moire potential is generated via the lattice constant mismatch between the bottom layer of pentalayer graphene and hBN, and a relative twist angle.
The lattice mismatch is $\epsilon_{G} = \frac{a_0}{a_{hBN}} - 1 \approx - 1.698\%$, where $a_0 = 0.246$nm and $a_{hBN} = 0.25025$nm are the respective lattice constants of monolayer graphene and hBN. 
The angle of rotation between the graphene and hBN later is taken to $0.77^{\circ}$, drawing inspiration from the experimental report \cite{lu2023fractional}.
The resulting \moire lattice  constant is $a_m = a_0 / \sqrt{\epsilon_{G}^2 + \theta^2} \approx 11.4$ nm.

\begin{figure}[t]
    \centering
\includegraphics[width=0.25\textwidth]{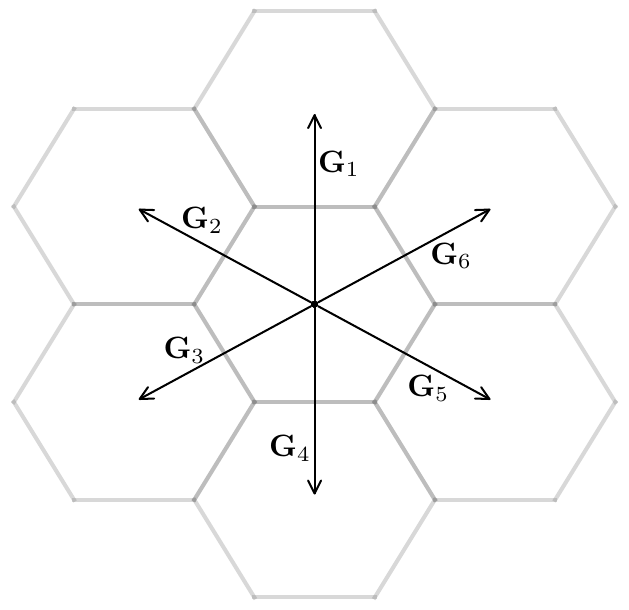}
    \caption{Moir\'e Reciprocal lattice vectors $\mathbf{G}_j$ \cite{jung_2018_prb}.}
    \label{fig_recip_vecs}
\end{figure}

The moir\'e Hamiltonian is defined as $ V_M(\mathbf{G})$ \cite{jung_2015_nc, jung_2018_prb}, where the \moire reciprocal lattice vectors are defined in Fig.~\ref{fig_recip_vecs} following the standard convention \cite{jung_2018_prb}.
Due to the two-sublattice nature of layer-1, we can define (for monolayer graphene-valley $K_G$) each $ V_M(\mathbf{G}_j)$ by a Pauli matrix in $\{A_j, B_j\}$ space, 
\begin{align}
    V_M(\mathbf{G_j}) = & \frac{ \sigma_0 + \sigma_z }{2} V_{AA} (\mathbf{G}_j) + \frac{ \sigma_0 - \sigma_z }{2} V_{BB} (\mathbf{G}_j) \nonumber \\ +
    & { \sigma_+ } V_{AB} (\mathbf{G}_j) +
    { \sigma_- }  \left[ V_{AB} \overline{\mathbf{G}}_j)\right]^* ,
\end{align}
where the final term arises from the fact that $H_M$ is Hermitian, and
\begin{align}
V_{AA/BB} (\mathbf{G}_{1,3,5}) &= \left[V_{AA/BB} (\mathbf{G}_{2,4,6})\right]^* = C_{AA/BB} e^{-i\phi_{AA/BB}} \\
    V_{AB}(\mathbf{G}_{1}) &= \left[V_{AB}(\mathbf{G}_{4})\right]^* = C_{AB} e^{2 \pi i /3} e^{-i \phi_{AB}} \\
    V_{AB}(\mathbf{G}_{5}) &= \left[V_{AB}(\mathbf{G}_{6})\right]^* = C_{AB} e^{-2 \pi i /3} e^{-i \phi_{AB}} \\
    V_{AB}(\mathbf{G}_{3}) &= \left[V_{AB}(\mathbf{G}_{2})\right]^* = C_{AB} e^{-i \phi_{AB}},
    \label{eq_moire_tunneling_terms}
\end{align}
where the ``rigid'' amplitudes and phases are presented in Table~\ref{tab:moire_params}.

For the valley polarization calculations, it is important to take into account the time-reversal partner of the \moire potential in Eq.~\ref{eq_moire_tunneling_terms}. 

\begin{table}[h!]
\centering
\begin{tabular}{|c|c|c|c|c|c|}
\hline
$C_{AA}$ [meV] & $C_{BB}$ [meV] & $C_{AB}$ [meV] & $\phi_{AA}$ & $\phi_{BB}$ & $\phi_{AB}$ \\
\hline
$-14.88$ & $12.09$ & $11.34$ & $50.19^{\circ}$ & $-46.64^{\circ}$ & $19.60^{\circ}$ \\
\hline
\end{tabular}
\caption{Moir\'e Hamiltonian parameters.}
\label{tab:moire_params}
\end{table}

Though the AA and BB terms are identical for both monolayer graphene valleys, the AB terms transform non-trivially under time-reversal (i.e. they are the complex conjugate of the $K_G$ monolayer valley term \cite{parkjeil2023}).

We note here that Ref. \onlinecite{lu2023fractional} presented a band structure with a Chern-$1$ band based on a phenomenological moire potential rather than a microscopic treatment.

\section{Methodologies: Hartree-Fock and Exact Diagonalization}

In this section, we provide details regarding the Hartree-Fock and Exact Diagonalization methods employed in this work.

\subsection{Hartree-Fock Framework}

The Hamiltonian can be written in terms of one-body piece $H_1 (\bfk) = H_0 (\bfk) + H_M(\bfk)$ (with the continuum kinetic term $H_0 (\bfk)$ given in Eq. \ref{full_hamiltonian_penta}, and \moire potential $H_M(\bfk)$ given in Eq. \ref{eq_moire_pot_main}), and two-body Coulomb interaction given in Eq. \ref{eq_screened_coulomb}.
These terms are written in the continuum plan-wave basis $c_{\mu}(\bfk)$, where $\mu$ denotes a generalized flavor degree of freedom.

Focusing on bands of interest within an energy window, it is beneficial to rewrite the Hamiltonian in the band basis, 
\begin{align}
    c( \bfk + \mathbf{G} ) = \sum_b \mu_{\mathbf{G}, b}(\mathbf{k}) \psi_b(\bfk),
\end{align}
where $\bfk$ is the momentum in the mBZ, $\mu_{{\v G}, b}(\bfk)$ is the Bloch function associated with band $b$ and reciprocal lattice vector $\mathbf{G}$, and $\psi_b(\bfk)$ is the fermionic band operator. 

The one-body term then simply becomes,
\begin{align}
H_1 (\bfk) = \sum_b \xi_b (\bfk) \psi_b ^{\dag} (\bfk) \psi_b(\bfk),
\end{align}
where $\xi_b (\bfk)$ is the dispersion (energy eigenvalue) of band $b$.

The two-body Coulomb interaction becomes,
\begin{widetext}
\begin{align}
H_C = \frac{1}{2A} \sum_{\bfk, \bfk', \bfq} \sum_{a,b,c,d} V^{\text{sc}}_C (\bfq) \lambda^{ab}(\bfk + \bfq, \bfk)  \lambda^{cd}(\bfk' - \bfq, \bfk') \psi_a ^{\dag} (\bfk+\bfq) \psi_c ^{\dag} (\bfk'-\bfq) \psi_d (\bfk') \psi_b (\bfk),
\end{align}
\end{widetext}
where $\{a,b,c,d\}$ denote band indices, and we have defined the form factors $\lambda^{ab}(\bfk + \bfq, \bfk) = \sum_{\mathbf{{G}}} \mu^* _{\mathbf{G},a}(\bfk+\bfq) \mu_{\mathbf{G},b}(\bfk)  $

Performing the mean-field decoupling, we obtain the following Hartree ($H_H$) and Fock ($H_F$) terms,

\begin{widetext}
\begin{align}
H_{H} = \frac{1}{A} \sum_{\bfk, \bfk'; \mathbf{G}; a,b,c,d} V(\mathbf{G})\lambda^{ab}(\bfk + \mathbf{G}, \bfk)  \lambda^{cd}(\bfk' - \mathbf{G}, \bfk') \rho_{cd}(\bfk') \psi_a^{\dag}(\bfk) \psi_b(\bfk),
\end{align}

\begin{align}
H_{F} = - \frac{1}{A} \sum_{\bfk, \bfk'; \mathbf{G}; a,b,c,d} V(\bfk - \bfk' +\mathbf{G})\lambda^{ab}(\bfk + \mathbf{G}, \bfk')  \lambda^{cd}(\bfk' - \mathbf{G}, \bfk) \rho_{cb}(\bfk') \psi_a^{\dag}(\bfk) \psi_d(\bfk),
\end{align}
    
\end{widetext}
where $\rho_{cd} (\bfk) = \langle \psi_c^{\dag} (\bfk) \psi_d (\bfk) \rangle$
is the mean-field amplitude to be determined self-consistently.

In performing the Hartree-Fock self-consistent calculations, we consider the conduction bands of interest, namely the `active' conduction band, and select a number of remote conduction bands within an energy window.
The occupied valence bands below the charge neutrality point are neglected for the large displacement fields. 
This neglect of the valence bands is aligned with the purpose of this work: to demonstrate the existence of FQAH states in a realistic setting. We leave detailed and accurate calculations for future studies.

\subsection{Exact diagonalization: projecting to active band}
To study fractional fillings, we perform exact diagonalization after projecting to the lowest Hartree-Fock band. Note that the mean-field effect of interaction $H_I$ is already taken into account in obtaining the Hartree-Fock band structure. To avoid double counting the interaction, we need to subtract the mean field contribution of interaction from the full interaction Hamiltonian before projection, namely
\begin{align}
    \tilde{H}_{ED} &= P_1 (H_1+H_C) P_1\\ \nonumber
    &= H^{MF}_K+P_1(H_C-H_H-H_F) P_1
\end{align}
where $H_1$ is the non-interacting piece of the Hamiltonian, $H^{MF}_K$ is the full mean field Hamiltonian which defines the mean field band structure, and $P_1=\sum_{\v k} \ket{\psi^1_{\v k}}\bra{\psi^1_{\v k}}$ is the projection operator to the first conduction band.
We note that we are fractionally filling the $\nu=1$ Hartree-Fock band in our ED study.

\textcolor{black}{
We note that this two-step calculation (of first computing the self-consistent Hartree-Fock bandstructure, and then projecting the total Hamiltonian to the lowest Hartree-Fock band) is  vital to our ED calculation, as it greatly reduces the Hilbert space of the fermions, thus enabling us to examine system sizes of up to 30 sites.
}

\section{Exact diagonalization data on other fractional fillings}
\label{app:exact}

In this section, we detail the findings of FQAH states at fillings $\nu=1/3, 2/3, 2/5, 
\text{and } 3/5$ for the spin-polarized model.

\begin{figure}
    \centering
    \vspace{10pt}
    \includegraphics[width=0.48\textwidth]{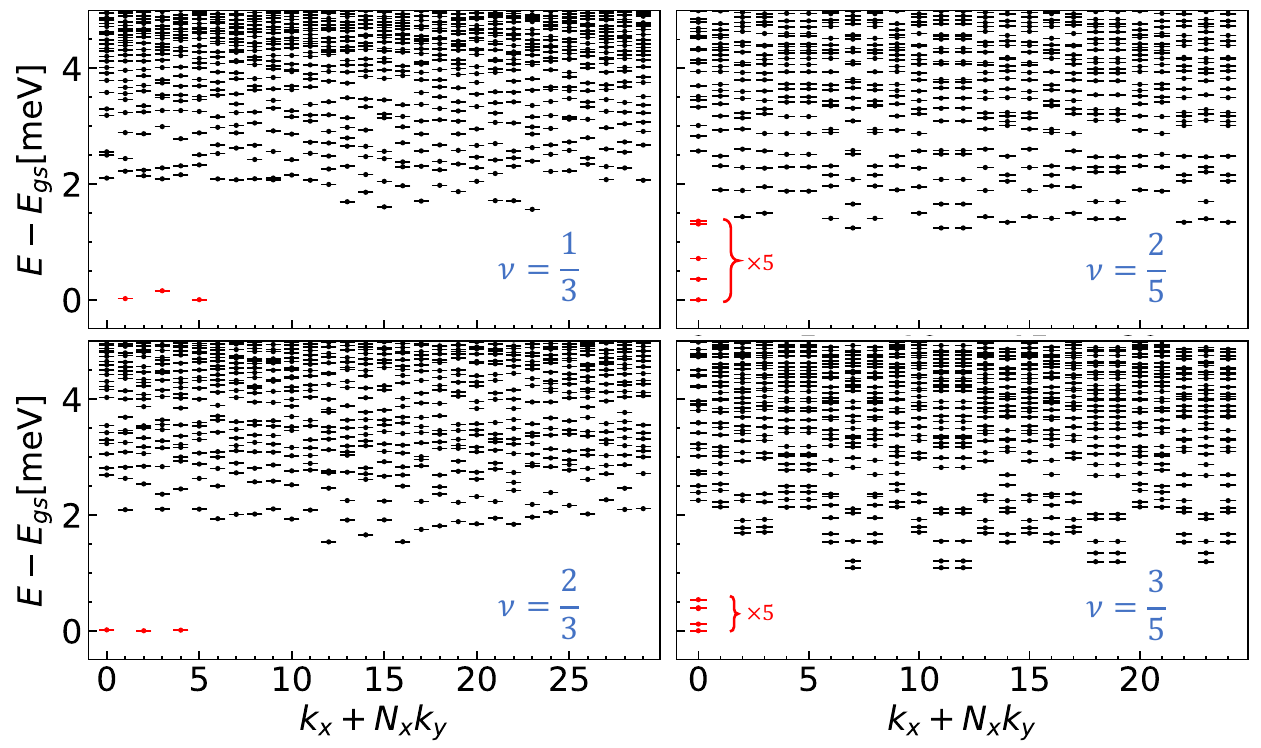}
    \caption{Low-energy spectrum of the spin-valley polarized system. (a)(c) $\nu=1/3$ and $\nu=2/3$, on a $N_x \times N_y = 6\times5$ system. (b)(d) $\nu=2/5$ and $\nu=3/5$ on a $5\times5$ system. These results are obtained for $\epsilon=8$, $D=-36$meV. }
    \label{spectrum}
\end{figure}

The low energy spectrums for these fillings are shown in Fig.~\ref{spectrum}. At all four fillings, we find approximate degeneracy consistent with FQAH states. The momentum spacing between these low-energy states is consistent with the expectation in Table.~\ref{tab:kshift}. The splitting of the five low-energy states of $2/5$ and $3/5$, respectively, is not as small as $1/3 \text{ and } 2/3$. We understand it as a consequence of a smaller system size of $5\times 5$ and the momentum matching between the five low-energy states of $\nu=2/5, 3/5$.

\begin{table}[h!]
\centering
\begin{tabular}{|c|c|c|c|}
\hline
Size & $\nu$ & $\Delta k_x/2\pi$ (mod 1) & $\Delta k_y/2\pi$ (mod 1) \\
\hline
$6\times5$ & 1/3 & 5/3 (2/3) & 2 (0)\\
$6\times5$ & 2/3 &-10/3 (2/3) & -4 (0) \\
$5\times5$ & 2/5 & 2 (0) & 2 (0) \\
$5\times5$ & 3/5 & -2 (0) & -2 (0) \\
\hline
\end{tabular}
\caption{Momentum shift (Eq.~3 in the main text) between topologically degenerate ground states. For $\nu>1/2$ FQAH states are mapped to $\nu'=1-\nu$ through a particle-hole conjugation. The minus sign in $\Delta k$ is due to the interchange of quasi-particle and quasi-hole.}
\label{tab:kshift}
\end{table}

\begin{figure}
    \centering
    \includegraphics[width=0.48\textwidth]{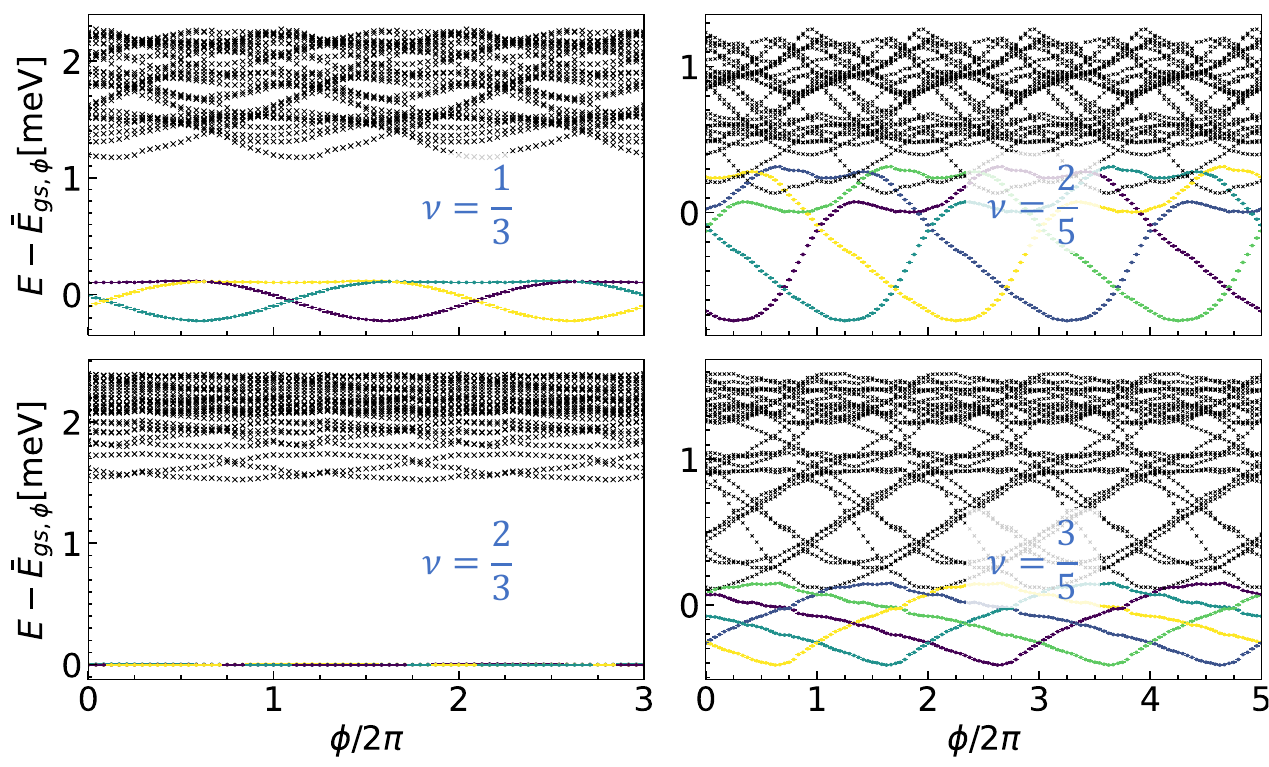}
    \caption{Spectral flow under flux threading along $x$ direction on (a)(c)  $6\times4$ system at $\nu=1/3$ and $2/3$. (b)(d) $5\times4$ system at $\nu=2/5$ and $3/5$. These results are obtained at $\epsilon=8$, $D=-36$meV. $\bar{E}_{gs}$ is the average energy for topologically degenerate ground states. The colors are identified by the momentum sector corresponding to the eigenstates.}
    \label{flow}
\end{figure}

In Fig.~\ref{flow} we demonstrate the spectrum flow under flux threading through one cycle of the torus. Under $2\pi$ flux threading along the $x$ direction, topologically degenerate ground states permute. For fillings $\nu=1/3, 2/3, \text{ and } 3/5$ the ground states are separated from excitations for all values of flux. For $\nu=2/5$ the topologically degenerate ground states mix with higher excitations for some values of flux. This is mainly due to the aforementioned small system size of $5\times4$, which splits the five degenerate ground states. Nonetheless, we are still able to separate these five states by tracking their momentum sectors (see colors in Fig.\ref{flow}), since excitation gaps in the same momentum sector never close during flux insertion.

\begin{figure*}
    \centering
    \includegraphics[width=0.8\textwidth]{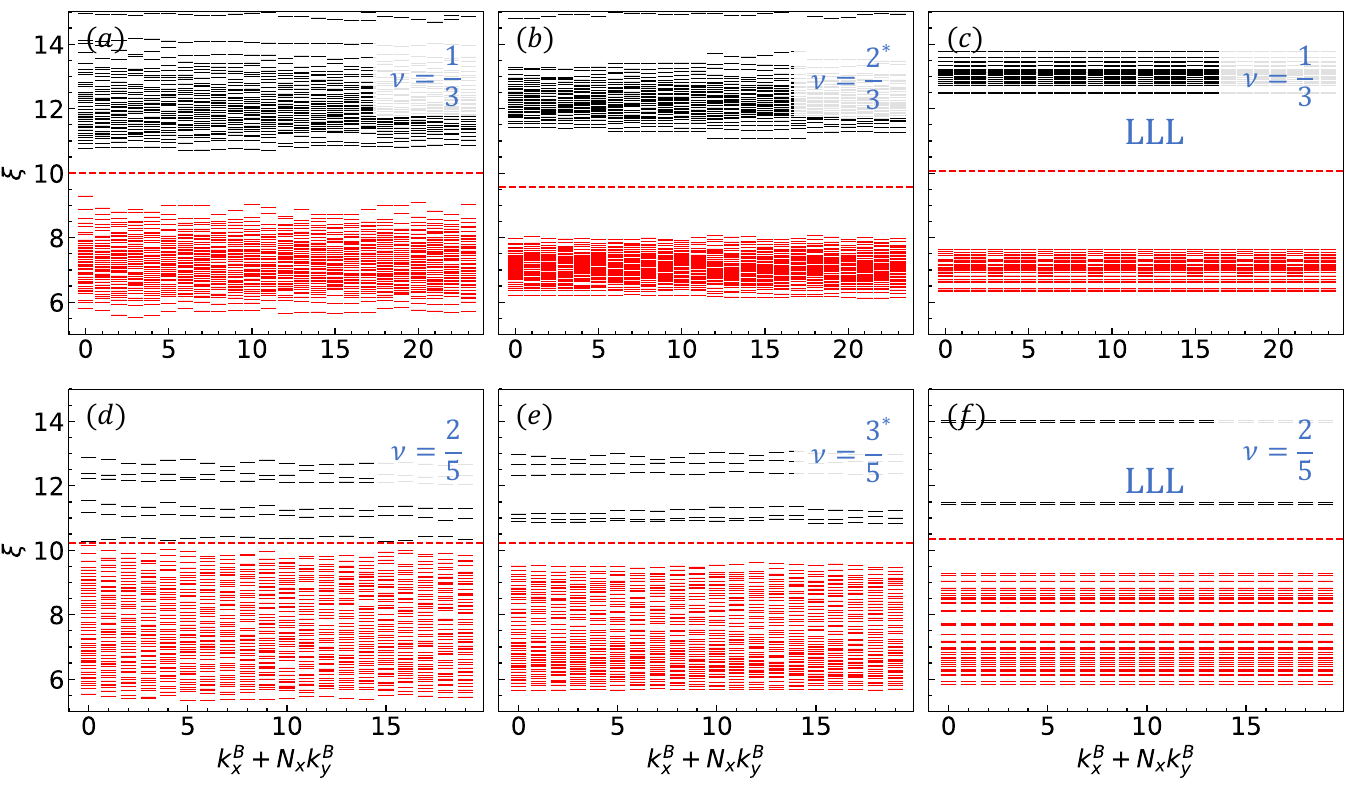}
    \caption{Particle entanglement spectrum of spin-polarized ground states and comparison to LLL. (a)(b)(c) $N_B=3$ on a $6\times4$ systems at filling $\nu=1/3$. The total number of states below the gap is 1088. (d)(e)(f) $N_B=3$ on a $5\times4$ system at filling $\nu=2/5$. The total number of states below the gap is 1020. The left and middle panels are PES for R5G/hBN with $\epsilon=8$, $D=-36$meV. (b) and (e) are computed for using particle-hole conjugate wavefunctions of $\nu=2/3$ and $3/5$. The right panels are PES for LLL with Coulomb interaction.}
    \label{PES}
\end{figure*}

To provide stronger evidence for FQAH and rule out the possibility of CDW states, we compute the particle entanglement spectrum (PES). In Fig.~\ref{PES} we present the PES for all four fillings alongside that of LLL with Coulomb interaction.

To compute PES, we construct the density matrix using the topologically degenerate ground states, for example at $\nu=1/3$, we consider the density matrix $\rho=\frac{1}{3}\left( \ket{\Psi_1}\bra{\Psi_1}+\ket{\Psi_2}\bra{\Psi_2}+\ket{\Psi_3}\bra{\Psi_3}\right)$. We then consider the particle bipartition of the $N$-electron system into two groups with $N_A$ and $N_B$ electrons. The reduced density matrix is obtained numerically by tracing over subsystem $A$, $\rho_B = \Tr_A{\rho}$. It retains the translation symmetry of the $N_x\times N_y$ lattice and therefore can be diagonalized in each momentum sector $k_B$. The `energy' is defined as $\xi_i =-\log\lambda_i$ where $\lambda_i$ are the spectrum of $\rho_B$. For fillings above $1/2$, we use the particle-hole conjugate wave function to construct the density matrix $\rho$. In Fig.~\ref{PES}(b) and (d) we denote the particle-hole conjugation by adding a `$^*$' on the filling fraction.

The PES is a hallmark of FQAH states because it reflects the Hilbert space dimension of low-energy quasi-hole excitations, which cannot be reproduced with a CDW state. In particular, for $\nu=1/3$ Laughlin state, the number of states below the PES gap is $Q=N_xN_y\frac{(N_xN_y-2N_B-1)!}{N!(N_xN_y-3N_B)!}$ \cite{regnault2011fractional}. With $N_x\times N_y = 6\times4$ and $N_B=3$, we observe $Q=1088$ states below the PES gap, consistent with the expectation. For fillings $\nu=2/5 \text{ and }3/5^*$, we show that for $N_x\times N_y = 5\times4$ and $N_B=3$, the number of states below the PES gap is $Q=1020$, the same as the counting found in LLL with Coulomb interaction.

We note that we find a robust FQAH state at $1/3$ filling, which is apparently not seen in experiments. A possible explanation is that, in our ED calculations, we used the Hartree-Fock bands obtained at a filling of $\nu = 1$ of the conduction band. As the filling is lowered the band structure may renormalize further and this may change the band parameters. While we expect this approximation to be a small effect for $\nu$ closer to $1$, at lower filling factors like $\nu = 1/3$, the change in the band parameters may affect the state. We leave a detailed study of this effect for the future.  

\textcolor{black}{
Finally, we note that the physical interpretation of our numerical results is that the FQAH states arise from hole doping the $\nu= 1$ state (either an anomalous Hall crystal or a moir\'e-enabled anomalous Hall crystal \cite{dong2024stability}), wherein the holes form the corresponding Jain states. 
This interpretation is rooted in the idea that the \moire potential pins the AHC crystal (or the \moire enabled anomalous Hall crystal \cite{dong2024stability}) with the corresponding commensuration energy gain. 
To that end, we posit that the major factor threatening the FQAH state is the destabilization of the underlying $\nu=1$ state. 
One may conjecture that the FQAH at fillings closer to $\nu=1$ will be more robust, say under the tuning of displacement fields, since the hole-doped AHC picture would be more relevant.
From an experimental point of view, the $\nu = 3/5$ state does indeed survive down to smaller displacement fields (as compared to the $\nu = 2/3$ state).
A smaller displacement field brings the electrons closer to the hBN layer, thus enhancing the moir\'e potential, which stabilizes the underlying AHC order and makes the hole-doped AHC picture more relevant at $\nu=3/5$ and smaller $D$ than at $\nu=2/3$ and larger $D$. This reasoning may explain why the FQAH state is more stable at $\nu=3/5$.
}

\section{Spin physics}
In contrast to the TMD system, in pentalayer graphene, the electron spin is potentially an active degree of freedom and is independent of the valley.  Thus we study the fate of the spin polarization, or lack thereof, in the FQAH states. 
Here we reconsider the role of the spin degree of freedom in R5G. As discussed\cite{zhou2021half,patri2023strong} for R3G, one approach is to assume that the interaction effects dominate over the \moire potential. Then it is possible that, even without a \moire potential,    spin polarization occurs at a high energy scale (leading, if valley polarization is also present, to a ``$1/4$-metal" where only one flavor degree of freedom is occupied) in the presence of interactions.  The \moire potential then serves to open up a band gap on the resulting spin and valley polarized bands. An alternate approach is to solve for the renormalized \moire band structure in the presence of valley polarization, and let the spin physics be determined by the remaining correlation effects. In R3G, the former approach was found\cite{patri2023strong} to be not quantitatively accurate on the topologically non-trivial side (and even qualitatively inaccurate for the topologically trivial side). For R5G, we have checked by exact diagonalization that at $\nu_T = 1$ the spin is fully polarized. For fractional fillings, we therefore undertook calculations of the many-body physics both for the case of a spin-polarized band and for the case where the spin physics is determined by correlation effects within the active conduction band. We present here the results without assuming spin polarization. We note that we take the Hartree-Fock band as a rigid band in which we populate electrons of either spin. This is an approximate treatment that should be refined in future work.

\label{app:spinful}
\begin{figure*}
    \centering
    \includegraphics[width=1\textwidth]{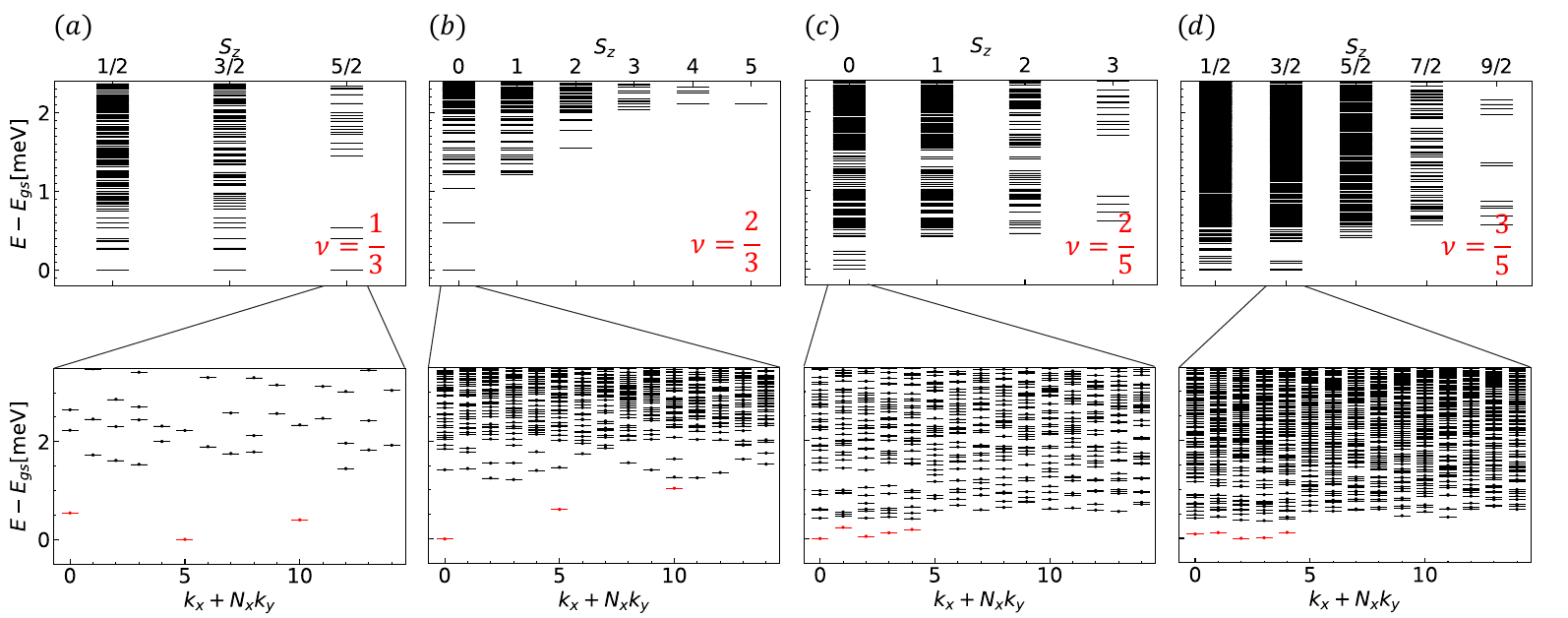}
    \caption{Low energy spectrum for the spinful model with a system size of $N_x\times N_y=5\times3$. The top panels are spectrum in all spin sectors. Due to SU(2) symmetry, the spectrum for a higher $S_z$ sector is always a subset of that of a lower $S_z$. Lower panels are momentum-resolved spectrum in the ground state $S_z$ sector (a) The ground states in all $S_z$ sectors are degenerate, indicating an SU(2) ferromagnet. (b)(c) Ground states are unpolarized. (d) Ground state is partially polarized, with $(\nu_\up, \nu_\down) = (2/5, 1/5)$.}
    \label{spinful_all}
\end{figure*}
In Fig.~\ref{spinful_all} we summarize the ED results including both spins. On the top row, we show the spectrum in all spin sectors. For filling $\nu=1/3$ the ground state is spin-polarized. For $\nu=2/3 \text{ and } 2/5$ the ground state is spin unpolarized. For $\nu=3/5$ we find the ground state is partially polarized. On the bottom row, we show the momentum-resolved spectrum within the ground state spin sector. Our results suggest the possibility of topologically degenerate ground states for all four fillings. These results point to Halperin-$(mnp)$ states: $(112)$ state at $\nu=2/3$, $(223)$ state at $\nu=2/5$. At $\nu=3/5$, $S_{tot}=3/2$ for 9 electrons, so the `layer' filling is $(\nu_\up, \nu_\down) = (2/5, 1/5)$, consistent with the $(321)$ state. We note that these results are obtained on a small system with $N_x\times N_y=5\times3$, and therefore the identification of topological order is not conclusive. 

In the lowest Landau level, the $2/3$ fractional quantum Hall state is well known to be spin unpolarized in the absence of Zeeman coupling. Similarly in previous calculations\cite{repellin2020chern} on models appropriate to the Chern band in twisted bilayer graphene aligned with a hBN substrate, both spin-unpolarized and spin-polarized states were found at this filling with a transition between them controlled by the details of the quantum geometry of the band. For the present system, we show our numerical ED data in Fig.~\ref{spinful_all}. At lattice filling $2/3$, we find that the ground states in different total $S_z$ sectors have different energies (with the lowest energy state living in the $S_z = 0$ sector). Thus this is a spin singlet state which we identify with the Halperin (112) state.  In contrast at $\nu = 1/3$, we find that the ground states in different total $S_z$ sectors are degenerate suggesting a spin-polarized state. We find a spin unpolarized state at $2/5$, and finally, at $\nu = 3/5$, a state that has partial spin polarization.

The results of this section emphasize the possibility of interesting spin physics of the FQAH state, As at the displacement fields of interest, the electrons mostly reside in the topmost layer, an in-plane magnetic field will have very little orbital coupling and will affect the physics solely through Zeeman coupling. Thus tilted field experiments should be a good experimental probe for spin physics.

\section{Hartree-Fock bands for $u_d = -40$ meV, and $\epsilon = 10$}

\begin{figure}
    \centering
    \vspace{5mm}
\includegraphics[width=0.5\textwidth]{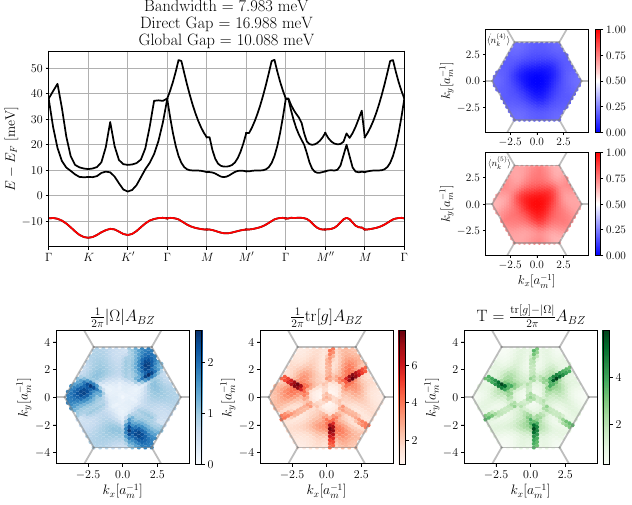}
    \caption{R5G/hBN Hartree-Fock (HF) band structure for $u_d=-40$meV, $\epsilon=10$, and gate distance $d_s=30$nm. 
    The HF calculation is performed with the conduction band (highlighted in red in the band structure) and two remote conduction bands (in black); the chemical potential is at 0. 
    The right-top subfigures show (for the active band) the layer-4,5 occupation number.
    The bottom subfigures are the Berry curvature, quantum metric, and trace condition. 
    The mean and variance of the trace condition over the BZ is $\langle T \rangle = 1.45$ and var$(T) = 0.97$, respectively, and taking into account the three lowest conduction bands.
    }
    \label{fig_hf_eps_10_ud_m40}
\end{figure}

We present in Fig.~\ref{fig_hf_eps_10_ud_m40} another representative plot of the Hartree-Fock R5G/hBN bandstructure and quantum geometric quantities for $u_d = -40$ meV and $\epsilon = 10$.
Qualitatively, they possess the same general features as the main text.

\begin{figure}[t]
    \centering
    \includegraphics[width=0.3\textwidth]{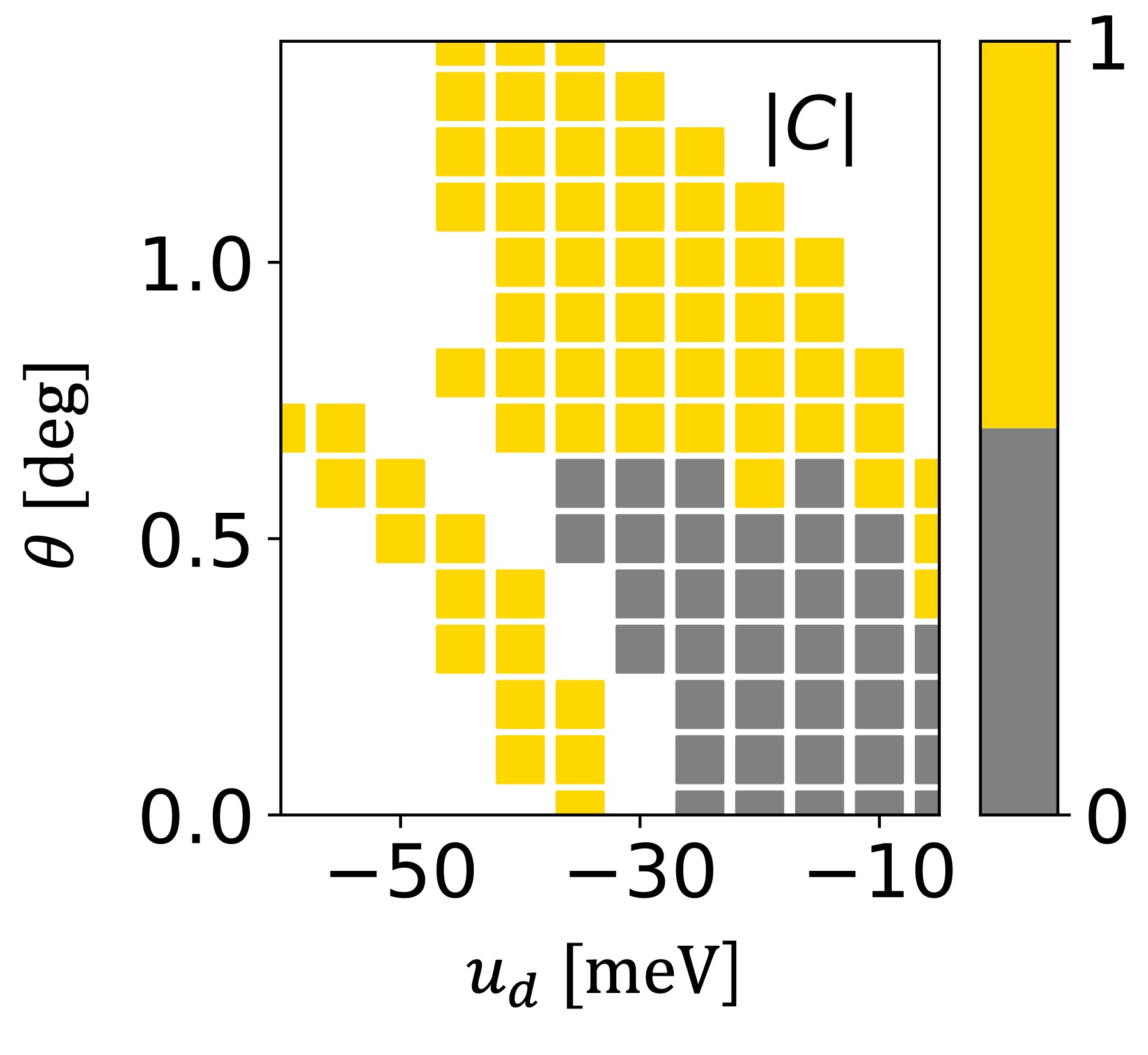}
    \caption{Twist angle ($\theta$) and displacement field energy ($u_d$) Hartree-Fock phase diagram without moir\'e potential for R5G. 
    The result is obtained for $\epsilon=6$ and without trigonal warping terms.
    }
    \label{fig_pd_nomoire}
\end{figure} 

\begin{figure*}
\includegraphics[width=0.465\textwidth]{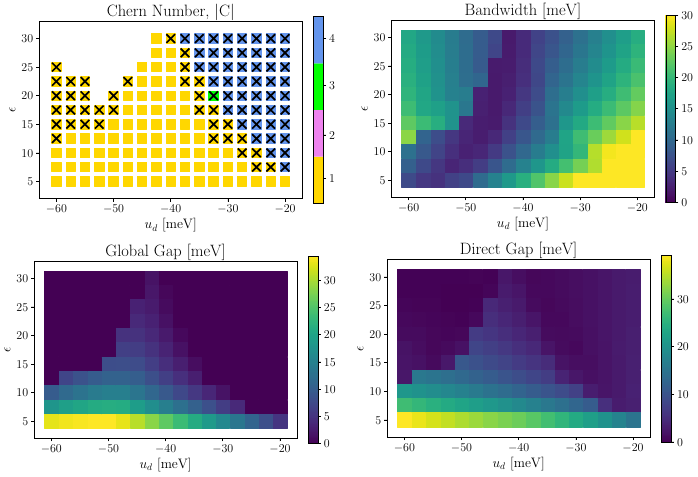}
\includegraphics[width=0.44\textwidth]{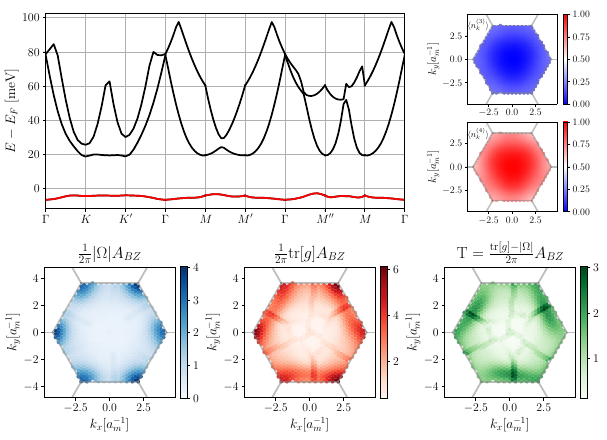}
    \caption{Left: R4G/hBN phase diagram indicating Chern number and bandwidth of the active band, global band gap, and direct band gap to nearest conduction band for different displacement field energies $u_d$ and dielectric constant $\epsilon$.
    The Chern number is well defined when the direct band gap is non-zero ($\geq0.5$meV).
    The `crossed' out yellow boxes in the Chern number indicate cases where the global bandgap is zero ($<0.5$meV), while the direct band gap is still non-zero (i.e. there is an indirect band gap -- the system is metallic).
    Phase diagrams are generated on a momentum mesh of 27$\times$27 and constructed using 19 \moire Brillouin zones, and using at least two distinct randomized initial mean-field ansatzes for a given parameter point.
    Right: A representative bandstructure and quantum geometric quantities are shown on the right for $u_d=-50$meV, $\epsilon=7.5$, and gate distance $d_s=30$nm. 
    The HF calculation is performed with the conduction band (highlighted in red in the band structure) and two remote conduction bands above; the chemical potential is at 0. 
    The right-top subfigures show (for the active band) the layer-3,4 occupation number.
    The bottom subfigures are the Berry curvature, quantum metric, and trace condition. 
    The mean and variance of the trace condition over the BZ is $\langle T \rangle = 1.1$ and var$(T) = 0.36$, respectively.
    Momentum mesh of 30 $\times$ 30 for this representative point, and taking into account the three lowest conduction bands.
    }
    \label{fig_4lg_hf_phase_diagram}
    \end{figure*}

\section{Moir\'e-less Hartree-Fock Phase Diagram}
\label{app_moireless_hf}

In this section, we consider the interaction effects on \moire-less R5G (i.e. zero \moire potential).
In addition, we have turned off the trigonal warping terms (i.e. retained only $\gamma_{0,1}$ in Eq. \ref{full_hamiltonian_penta}) to solely focus on the role of interactions in a minimal setting. 
Figure \ref{fig_pd_nomoire} is the Hartree-Fock phase diagram with zero moir\'e potential. 
The result is qualitatively the same as the full model of Eq. \ref{full_hamiltonian_penta}, namely a $C=1$ Chern insulator is still found over a wide range of displacement field at the experimentally relevant twist angle $\theta=0.77^\circ$. Our results are in agreement with those reported in Refs. \cite{dong2023anomalous,zhou2023fractional}

There are a few salient features from Fig. \ref{fig_pd_nomoire}.
Firstly, the qualitatively similar behavior of the simplified model as compared to the full model suggests that the interaction-induced Chern insulator does not rely much on the details of the model.
This suggests a universal mechanism for interactions to favor Chern insulators.
We leave the discussion of this universal mechanism to future work. 
This state, where spontaneous breaking of continuous translation symmetry opens a single particle gap and the filled Hartree-Fock bands have a net Chern number, is known as a `Hall crystal'\cite{PhysRevB.39.8525}. Thus the Hartree-Fock calculation of the full model can be interpreted as describing a Hall crystal that is pinned by the \moire potential. 
However, it is important to note that the Hartree-Fock treatment is strongly biased towards finding a translational symmetry breaking mean-field solution. We expect that it is safer to use Hartree-Fock in the presence of the \moire potential than without. 
Therefore, the question of the possibility of a Hall crystal phase in R5G with misaligned hBN (i.e. zero \moire potential) needs to be addressed with more care, for example, through a numerical study beyond Hartree-Fock.

\section{R4G/hBN Hartree-Fock Phase diagram}
\label{r4g}
In this section, we consider the HF-phase diagram for R4G/hBN under the influence of displacement field and dielectric screening.
The continuum dispersion of R4G is modelled by the Hamiltonian\cite{parkjeil2023}, 
\begin{widetext}
\begin{align}
\label{full_hamiltonian_tetra}
H_{R4G} = 
\begin{pmatrix}
\frac{3u_d}{2} & v_0^{\dag} & v_4^{\dag} & v_3 & 0 & \frac{\gamma_2}{2} & 0 & 0 \\
v_0 & \frac{3u_d}{2} + \delta & \gamma_1 & v_4^{\dag} & 0 & 0 & 0 & 0  \\
v_4 & \gamma_1 & \frac{u_d}{2} + u_a & v_0^\dag & v_4^\dag & v_3 & 0 & \frac{\gamma_2}{2} \\
v_3^{\dag} & v_4 & v_0 & \frac{u_d}{2} + u_a & \gamma_1 & v_4^{\dag} & 0 & 0  \\
0 & 0 & v_4 & \gamma_1 & -\frac{u_d}{2} + u_a & v_0^{\dag} & v_4^{\dag} & v_3 \\
\frac{\gamma_2}{2} & 0 & v_3^{\dag} & v_4  & v_0  & -\frac{u_d}{2} u_a & \gamma_1 & v_4 ^{\dag}  \\
0 & 0 & 0 & 0 & v_4 & \gamma_1 & -\frac{3u_d}{2} + u_a & v_0^{\dag}  \\
0 &0 & \frac{\gamma_2}{2} & 0 & v_3^{\dag} & v_4 & v_0 & -\frac{3u_d}{2}
\end{pmatrix},
\end{align}
\end{widetext}

where we use the same tight-binding parameters of R5G (listed in from Tab. \ref{tab:tight_binding}). 
Just as with R5G/hBN, the displacement field energy drop between adjacent layers is $u_d$.
However, a distinction is that the potential difference between the top and bottom layers is $3 u_d$ for R4G/hBN, rather than $4 u_d$ for R5G/hBN (the reason is due to the even/odd number of layers for four-layer/five-layer systems, as seen in Eq. \ref{full_hamiltonian_tetra}).
As such, when comparing the different platforms, it is instructive to consider $ u_{TB} \equiv u_T - u_B$.

We present the `phase diagram' for the R4G in Fig.~\ref{fig_4lg_hf_phase_diagram}.
The general features of the R5G/hBN phase diagram translate to R4G/hBN. While there is a wide region of $|C|=1$ in parameter space, however, there is a restricted region suitable for FQAH states (namely, a flat band, with a robust gap to neighboring bands).
The optimal displacement field energy is $|u_d| \approx $50 meV.
This corresponds to $u_{TB} ^{(4)} \approx $150 meV, while for R5G/hBN,  $u_{TB} ^{(5)} \approx $140 meV (taking an optimal displacement field energy of 35 meV).
We present a representative `optimal' Hartree-Fock band structure in Fig.~\ref{fig_4lg_hf_phase_diagram}.
As seen, the band is flat with bandwidth $\approx 4$meV, and well separated from the neighboring conduction bands (direct gap of $\approx$ 23 meV).
The bands have an almost $C_6$ symmetry, a feature reflected in the accompanying Berry curvature.

\begin{figure}[t]    \includegraphics[width=0.45\textwidth]{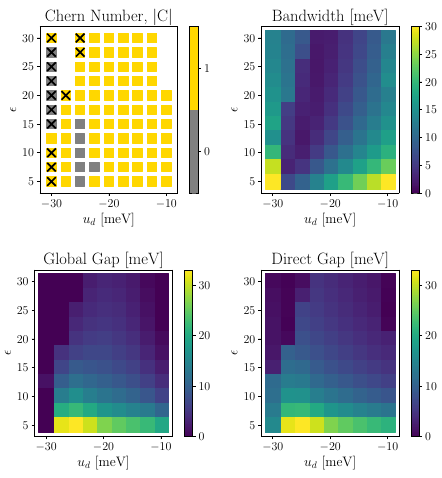}
    \caption{R6G/hBN phase diagram indicating Chern number and bandwidth of the active band, global band gap, and direct band gap to nearest conduction band for different displacement field energies $u_d$ and dielectric constant $\epsilon$.
    The Chern number is well defined when the direct band gap is non-zero ($\geq0.5$meV).
    The `crossed' out yellow boxes in the Chern number indicate cases where the global bandgap is zero ($<0.5$meV), while the direct band gap is still non-zero (i.e. there is an indirect band gap -- the system is metallic).
    Phase diagrams are generated on a mesh of 24$\times$24 and constructed using 19 \moire Brillouin zones, and using at least two distinct randomized initial mean-field ansatzes for parameter points, and taking into account the four lowest conduction bands.
    }
    \label{fig_6lg_hf_phase_diagram}
    \end{figure}
    
\section{R6G/hBN Hartree-Fock Phase diagram}
\label{r6g} 

In this section, we consider the HF-phase diagram for R6G/hBN under the influence of displacement field and dielectric screening.
The continuum dispersion of R6G is adapted from the R4G and R5G models in this work and given in Eq. \ref{full_hamiltonian_hepta}.
We note that we use the same tight-binding parameters of R5G (listed in from Tab. \ref{tab:tight_binding}).
We depict in Fig.~\ref{fig_6lg_hf_phase_diagram} the Hartree-Fock phase diagram over a narrower range of displacement field than the R5G/hBN.
In particular, a suitable window to examine FQAH states lies in the range of interlayer displacement field energy of $\approx -20$meV.

\setcounter{MaxMatrixCols}{20}
\begin{widetext}
\begin{align}
\label{full_hamiltonian_hepta}
H_{R6G} = 
\begin{pmatrix}
\frac{5u_d}{2}  & v_0^{\dag} & v_4^{\dag} & v_3 & 0 & \frac{\gamma_2}{2} & 0 & 0 & 0 & 0 & 0 & 0 \\
v_0 & \frac{5u_d}{2} + \delta & \gamma_1 & v_4^{\dag} & 0 & 0 & 0 & 0 & 0 & 0 & 0 & 0 \\
v_4 & \gamma_1 & \frac{3u_d}{2} + u_a & v_0^\dag & v_4^\dag & v_3 & 0 & \frac{\gamma_2}{2} & 0 & 0 & 0 & 0 \\
v_3^{\dag} & v_4 & v_0 & \frac{3u_d}{2} + u_a & \gamma_1 & v_4^{\dag} & 0 & 0 & 0 & 0 & 0 & 0 \\
0 & 0 & v_4 & \gamma_1 & \frac{u_d}{2} + u_a & v_0^{\dag} & v_4^{\dag} & v_3 & 0 & \frac{\gamma_2}{2} & 0 & 0 \\
\frac{\gamma_2}{2} & 0 & v_3^{\dag} & v_4  & v_0  & \frac{u_d}{2} + u_a & \gamma_1 & v_4 ^{\dag} & 0 & 0 & 0 & 0 \\
0 & 0 & 0 & 0 & v_4 & \gamma_1 & \frac{-u_d}{2} + u_a & v_0^{\dag} & v_4^{\dag} & v_3 & 0 & \frac{\gamma_2}{2}  \\
0 & 0 & \frac{\gamma_2}{2} & 0 & v_3^{\dag} & v_4 & v_0 & \frac{-u_d}{2} + u_a & \gamma_1 & v_4^{\dag} & 0 & 0 \\
0 & 0 & 0 & 0 & 0 & 0 & v_4 & \gamma_1 & \frac{-3u_d}{2} + u_a & v_0^{\dag} & v_4^{\dag} & v_3 \\
0 & 0 & 0 & 0 & \frac{\gamma_2}{2} & 0 & v_3^{\dag} & v_4 & v_0 & \frac{-u_d}{2} + u_a  & \gamma_1 & v_4^{\dag} \\
0 & 0 & 0 & 0 & 0 & 0 & 0 & 0 & v_4 & \gamma_1 & \frac{-5u_d}{2} + \delta & v_0^{\dag} \\
0 & 0 & 0 & 0 & 0 & 0 & \frac{\gamma_2}{2} & 0 & v_3^{\dag} & v_4 & v_0^{\dag} & \frac{-5u_d}{2}  \\
\end{pmatrix}.
\end{align}
\end{widetext}

\end{document}